\documentclass[twocolumn,showpacs,preprintnumbers,amsmath,amssymb,aps,prb,superscriptaddress]{revtex4-1}

\pdfoutput=1
\usepackage[pdftex]{graphicx}
\usepackage[english]{babel}
\usepackage{soul}
\usepackage{cleveref}

\usepackage{dcolumn}
\usepackage{bm}

\usepackage{color}
\usepackage{amstext}
\usepackage{braket}

\begin{document}


\title{Multi-hole edge states in SSH chains with interactions}

\author{A. M. Marques}
\email{anselmomagalhaes@ua.pt}
\affiliation{Department of Physics $\&$ I3N, University of Aveiro, 3810-193 Aveiro, Portugal}
\author{ R. G. Dias}
\affiliation{Department of Physics $\&$ I3N, University of Aveiro, 3810-193 Aveiro, Portugal}

\date{\today}


\begin{abstract}
We address the effect of nearest-neighbor (NN) interactions on the topological properties of the Su-Schrieffer-Heeger (SSH) chain, with alternating hopping amplitudes $t_1$ and $t_2$.
Both numerically and analytically, we show that the presence of interactions induces phase transitions between topologically different regimes. 
In the particular case of one-hole excitations in a half-filled SSH chain,
the $V/t_2$ vs. $t_1/t_2 $ phase diagram has topological phases at diagonal regions of the phase plane. The interaction acts in this case as a passivation potential.
For general filling of the SSH chain, different eigensubspaces of the SSH Hamiltonian may be classified as topologically trivial and non-trivial. The two-hole case is studied in detail in the large interaction limit, and we show that a mapping can be constructed of the two-hole SSH eigensubspaces into one-particle states of a non-interacting one-dimensional (1D) tight-binding  model, with  interfaces between regions with different hopping constants and local potentials. 
The presence of edge states of topological origin in the equivalent chain can be readily identified, as well as their correspondence to the original two-hole states. Of these states only some, identified by us, are protected and, therefore, truly topological. Furthermore, we found that the presence of the NN interaction generates a state where two holes occupy two consecutive edge states.
Such many-body states should also occur for arbitrary filling leading to the possibility of a  macroscopic hole gathering at the surface (at consecutive edge states). 
\end{abstract}


\maketitle

\section{Introduction}
\label{sec:intro}

The SSH model of polyacetilene chains \cite{Su1979} is an extensively studied 1D tight-binding model with alternate hopping constants, which can support topologically protected edge states \cite{Asboth2016}.
When interactions are introduced in the SSH model, the independent electron picture no longer holds and one can not determine a band structure from which the Berry phase could be calculated. 
However, as we show below, in a half-filled SSH chain (one electron per site) with NN interactions, one-hole excitations can be treated as independent effective particles, and one recovers the Berry phase.
The effect of on-site interactions on the topological excitations of a half-filled 1D chain was already characterized numerically by Guo and Shen \cite{Guo2011}.

The possibility of topological phases in real materials taking into account the presence of many-body interactions has been first addressed by Niu and Thouless \cite{Niu1984}. 
By interpreting a 1D chain as a cell of a larger supercell of equivalent chains, one can employ the method of twisted boundary conditions on the many-body wave function to find, after averaging over all possible boundary conditions, a quantized Berry phase \cite{Xiao2010}. 
The method works as long as the system remains an insulator as interactions are introduced.
This approach was recently followed in the case of interacting 1D chains with fractional fillings where, due to the degeneracy of the ground state, a topological phase characterized by a fractional Berry phase was found \cite{Guo2012,Budich2013}.

In the context of SSH chains, different kinds of interactions have been introduced and their effects characterized: Hubbard interaction \cite{Kivelson1982}, impurity atoms at specific sites \cite{Glick1986,Glick1988,Rossi1992}, spin-orbit coupling \cite{Yan2014}, superconducting pairing terms \cite{Sticlet2014} and electron-electron (e-e) interactions between nearest neighbors in periodic chains \cite{Weber2015}.
Hubbard and e-e interactions have also been studied in the context of polaron transport dynamics \cite{Ma2009,Cunha2014,Zhang2016}. 
Topologically protected edge states in similar 1D optical lattices with on-site interactions \cite{Grusdt2013}, or with Zeeman \cite{Liu2013} and synthetic gauge fields \cite{Li2015}, have also been studied.
The problem of two-body physics in 1D chains \cite{Valiente2008,Javanainen2010,Wang2010,Nguenang2012,Qin2014} has attracted the attention of the community as of lately and, in the particular context of the SSH model, which is the focus of this paper, there are some very recent papers that address the problem of two-boson states (doublons) in the presence of Hubbard interactions \cite{Bello2016,Liberto2016,Gorlach2016}.

In this paper, we consider interactions in a fermionic half-filled SSH chain and show how, for one-hole excitations, the NN interactions are converted into local potentials at the edge sites which, at critical strengths, reverse the topological nature of the chain \cite{Rossi1992,Bello2016,Liberto2016}. 
We further study two-hole excitations and find that, in the limit of strong interactions, the two-hole states available in a given eigensubspace can be translated as one-particle states in an equivalent chain with different sections, whose construction rules we detail, where the interaction vanishes. Because of this, in all different sections of this equivalent chain the usual topological characterization, given by the Berry phase, can again be made, and thus the possible presence of topological states can be readily identified.

\section{The model}
\label{sec:the_model}

\begin{figure}[h]
\begin{center}
\includegraphics[width=0.47 \textwidth]{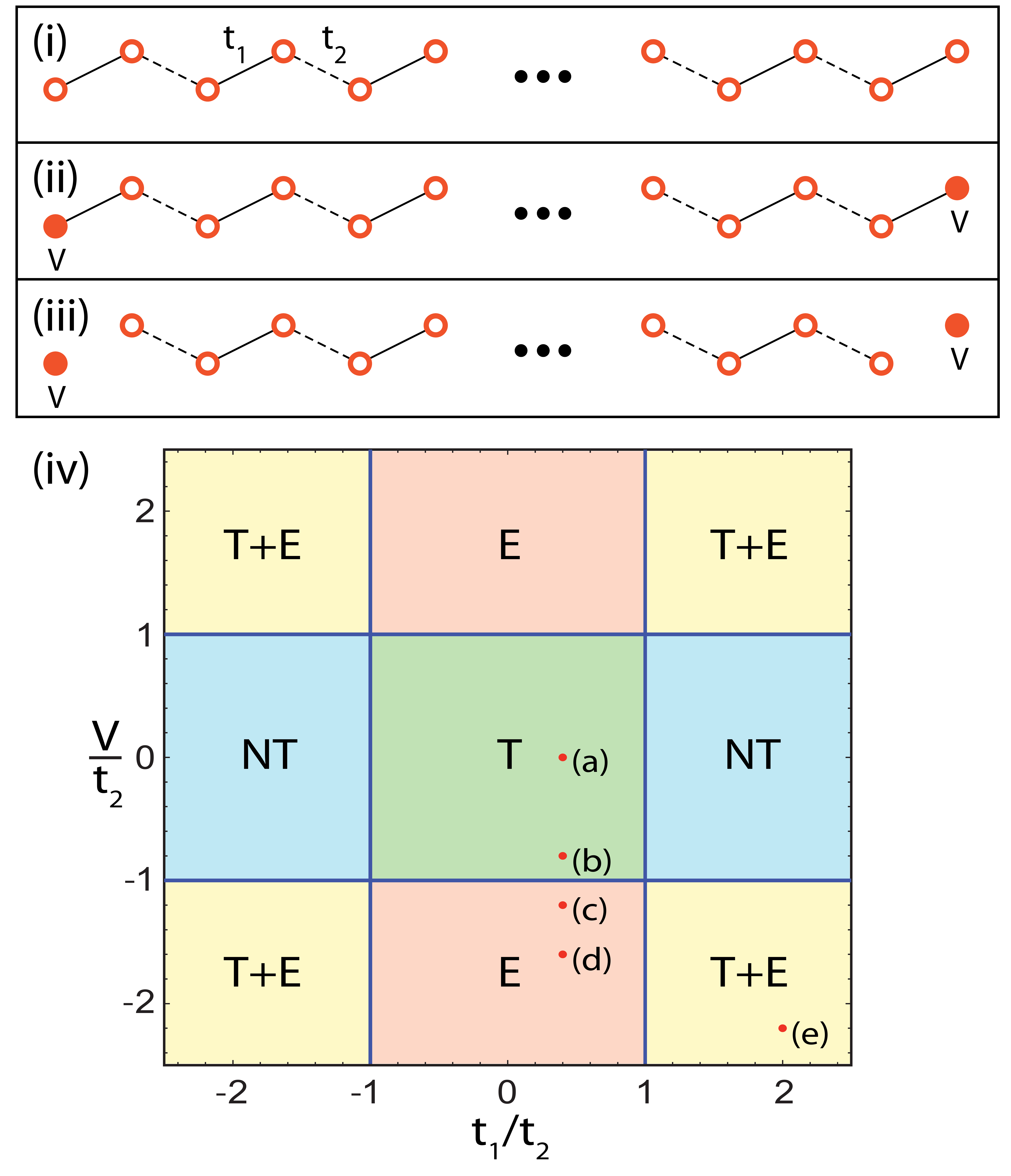}
\end{center}
\caption{(i) Open SSH chain with alternating $t_1$ and $t_2$ hoppings, represented by solid and dashed lines respectively. (ii) Added on-site potentials at the edges. 
(iii) When $V\to \infty$ the system can be splitted in two independent parts: the two edge sites and the rest of the chain. The edge hoppings of this smaller chain switch from $t_1$ to $t_2$, so that its topological nature, for fixed $(t_1,t_2)$, is the  inverse of (i).
(iv) $V/t_2$ vs. $t_1/t_2$ phase diagram of a one-hole state in an infinite open chain. 
In region T, there are midgap topological edge states. In regions NT, there are no edge states. 
In regions E, there are non-topological edge states pinned around the impurities with energies above the conduction bands (for $V>t_2$) or below the valence bands (for $V<-t_2$). 
In regions T+E, there are both topological and non-topological edge states. Points labeled (a)-(e) show the position on the phase diagram of the corresponding cases in Fig.~\ref{fig:energylevels}.}
\label{fig:ssh}
\end{figure}

We consider a spinless SSH model of an open chain with interactions, depicted in Fig.~\ref{fig:ssh}(i),
\begin{eqnarray}
\label{eq:hamiltonian}
H=&-&t_1\sum_{j=1}^{N/2}\big(c^{\dagger}_{2j-1}c_{2j} +H.c.\big) \nonumber
\\
&-&t_2\sum_{j=1}^{N/2-1} \big(c^{\dagger}_{2j}c_{2j+1} +H.c.\big) + V\sum_{j=1}^{N-1} n_{j}n_{j+1},
\end{eqnarray}
where $j$ is the site number, $n_{j}=c^{\dagger}_{j}c_{j}$ is the electronic occupation number, $N$ is the number of sites, $t_1$ and $t_2$ are the staggered hopping parameters of the chain and $V$ is the NN Coulomb interaction.

\subsection{One-hole states}

Let us consider one-hole excitations in an half-filled chain (one electron per site). 
Using the particle-hole transformation $c^{\dagger}_j(c_j)\to h_j(h^{\dagger}_j)$, together with the fermionic anticommutation rules, the Hamiltonian in (\ref{eq:hamiltonian}) becomes [apart from a constant term given by $V(N-1-2)$, where $N-1$ is the number of links and 2 is the number of missing links for a hole placed at a bulk site],
\begin{eqnarray}
\label{eq:hamiltonian_hole}
H&=&t_1\sum_{j=1}^{N/2}\big(h^{\dagger}_{2j-1}h_{2j} +h.c.\big)  \nonumber
\\
&+&t_2\sum_{j=1}^{N/2-1}\big(h^{\dagger}_{2j}h_{2j+1} +h.c.\big) + V \big(n_1^h+n_{N}^h\big),
\end{eqnarray}
where $n_j^h=h^{\dagger}_jh_j$ is the hole occupation number. The last term shows how the NN interaction translates into an on-site potential at the end sites, that is, in the one-hole Hamiltonian the interaction becomes equivalent to an impurity (passivation) potential located at both ends, reflecting the different coordination number of the end sites, as shown in Fig.~\ref{fig:ssh}(ii). 

This edge potential $V$ will change the relative position of the energy of the topological edge states and, with increasing $V$ (more precisely for $\vert V \vert > t_2$), new non-topological localized states appear, pinned at the edge impurities \cite{Rossi1992,Bello2016,Liberto2016}, and the topological nature of the chain is reversed, \textit{i.e.}, it goes from topologically non-trivial to trivial, or vice-versa. 
The phase diagram of Fig.~\ref{fig:ssh}(iv) illustrates this behavior. 
The $V=0$ line gives the usual SSH topological characterization. 

There is a simple qualitative picture to show why a strong $\vert V \vert$ changes the topological nature of the chain. 
In the absence of impurity potentials at the edges, topological edge states are present if the edge $t_1$ hoppings  are the long bonds ($t_1<t_2$), and absent otherwise. 
If we consider the $\vert V \vert \to \infty$ limit, the chain can be divided in two independent parts \cite{Glick1986}: the edge sites separate from the rest of the chain, as shown in Fig.~\ref{fig:ssh}(iii). 
The edge hoppings of the new smaller chain switch from $t_1$ to $t_2$, so the topologically non-trivial phase requires the $t_2$ hoppings to be now the long bonds - the dimerization of the chain is reversed \cite{Rossi1992,Wada1992}.
Numerically it is found that an effective dimerization reversal occurs at $\vert V \vert=t_2$, which is also the $\vert V \vert$ value above which localized impurity states appear (a similar description can be found in section 2 of the appendix in Ref.~\onlinecite{Liberto2016}).

\begin{figure}[t]
\begin{center}
\includegraphics[width=0.47 \textwidth]{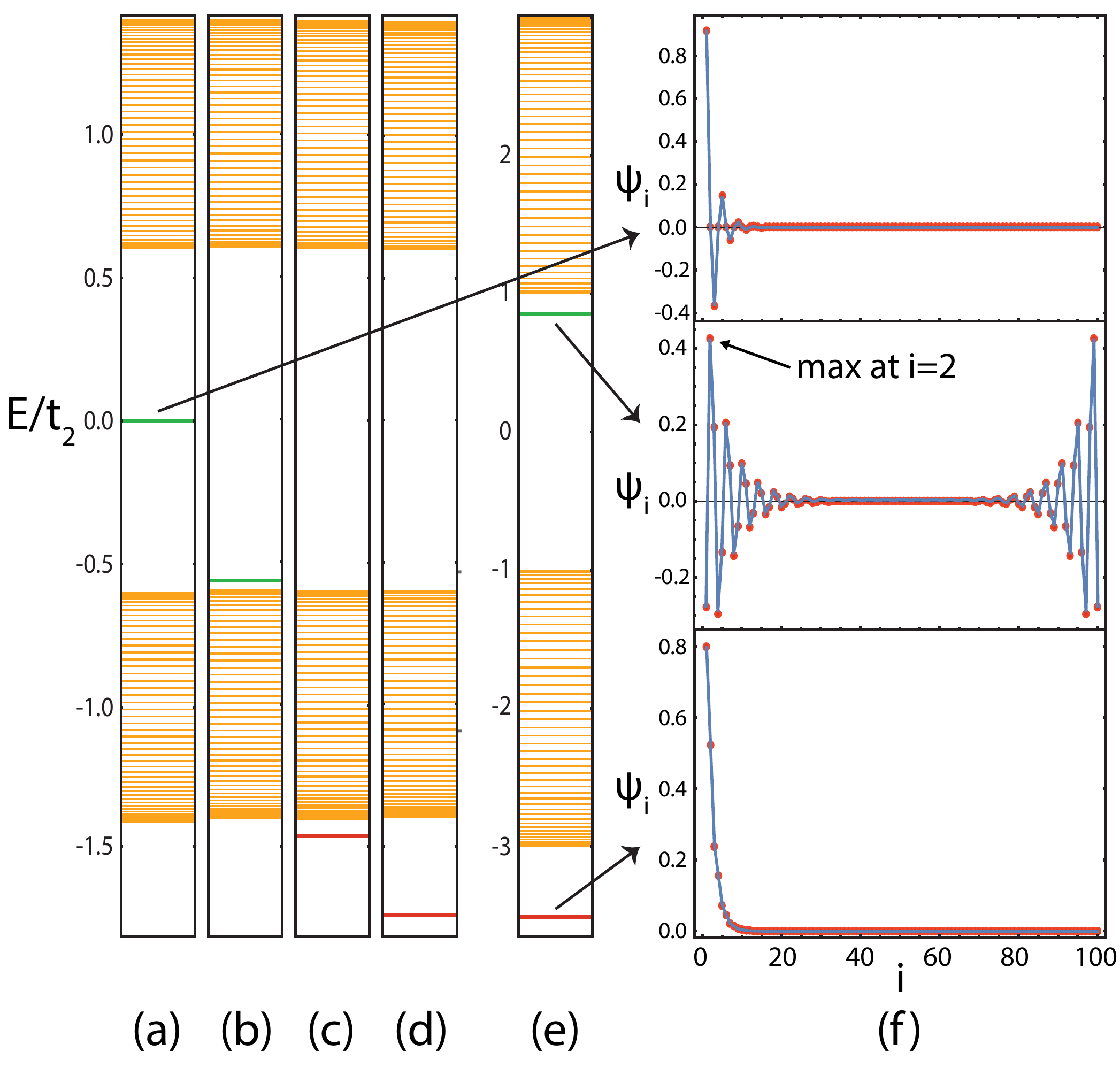}
\end{center}
\caption{Energy spectrum for one-hole states of a chain with $N=100$ sites for $t_1=0.4$ and (a) $V=0$, (b) $V=-0.8$, (c) $V=-1.2$, (d) $V=-1.6$, and (e) $t_1=2.0$ and $V=-2.2$, in units of $t_2$ [see corresponding labeled points in Fig.~\ref{fig:ssh}(iv)]. Green levels correspond to topological edge states and red levels to the impurity edge states. (f) Normalized probability amplitude distribution on the sites of the chain of the edge states indicated by the arrows (edge states are doubly degenerate, only one is shown here for each case).}
\label{fig:energylevels}
\end{figure}

A concrete example of a chain with $N=100$ sites was studied. The phase diagram of Fig.~\ref{fig:ssh}(iv) was seen to hold for this chain, apart from small finite-size effects. 
For different combinations of parameters, in different regions of the phase diagram, the energy levels are plotted in Figs.~\ref{fig:energylevels}(a)-(e). 
We considered only negative values for $V$. 
The energy levels for the symmetric positive $V$'s are given by a reflection about the zero energy level.
A finite detachment of an energy level from either of the bands was the criterion used for defining the edge states.

The usual energy spectrum for an SSH chain in the topological phase is recovered when $V=0$, as in Fig.~\ref{fig:energylevels}(a), with its zero energy states (doubly degenerate green level).
By turning on a small $V$, the only effect will be to lower the topological edge states close to the top of the lower band [see Fig.~\ref{fig:energylevels}(b)].
A further increase in $V$ and the aforementioned topological transition takes place: the topological edge state disappears as it merges with the lower band and, at the same time, impurity edge states appear below the lower band, as becomes clear by contrasting Figs.~\ref{fig:energylevels}(b) and (c), respectively a little before and after the topological transition. 
Continuing to increase $V$ only lowers the energy of the impurity edge states, as in Fig.~\ref{fig:energylevels}(d).
If, one the other hand, the set of parameters falls in one of the yellow regions in the phase diagram [see point (e) in Fig.~\ref{fig:ssh}(iv)], as is the case of Fig.\ref{fig:energylevels}(e), both topological and impurity edge states are present. 
These two kinds of edge states appear simultaneously at the transition from a non-topological blue region to a yellow region in Fig.~\ref{fig:ssh}(iv).

Examples of the spatial distribution of the wavefunction of some of the edge states in the chain are shown in Fig.~\ref{fig:energylevels}(f). 
The bottom case is of an impurity edge state, which is also the ground state.
Both the top and middle cases concern topological edge states.
The most important difference between them is that the maximum in the probability amplitude occurs at an edge site in the top case, and at the first inner sites at both edges (i=2,$N-1$) in the middle case. 
This comes as a consequence of the dimerization reversal which occurs in the middle case of Fig.~\ref{fig:energylevels}(f), where $\vert V \vert>t_2$. 
In the $\vert V \vert \to \infty$ limit illustrated in Fig.~\ref{fig:ssh}(c), the wavefunction of the bottom case of Fig.~\ref{fig:energylevels}(f) would be completely localized at the edge site and, in the middle case, the probability amplitude at the edge sites would be zero, since the edge sites would become  independent of the chain.

The results derived here for one-hole states can be used to describe other kinds of one-particle states.
Consider the introduction of an electron into an half-filled extended-Hubbard chain \cite{Hirsch1982,Cannon1991,Voit1992,Mila1993,Dongen1994,Penc1994,Ejima2016} of spinfull electrons for $U\gg V\gg t_1,t_2$. 
This condition ensures that the background is composed of one electron per site.
The introduction of another electron into the chain creates a doubly occupancy. 
The effect of the e-e interaction is again dependent on the position of the double occupancy, that is, on the number of nearest neighbors: if it is located at a bulk site, the energy is raised by $2V$ and if, on the other hand, it is located at an edge site the energy is raised by $V$ (the effect is symmetric to the one-hole states considered before).
Note that both the spin configuration of the background and of the introduced electron become irrelevant.
Then, after the constant $U$ term, giving the energy of the double occupancy, is taken out, the problem can be treated, as before, as the creation of a one-hole state in a background of spinless electrons, with the introduction of an on-site potential $-V$ at the edges [the same as Fig.~\ref{fig:ssh}(ii), but with $V\to -V$].

Another example is the quantum Heisenberg XXZ model \cite{Orbach1958} with staggered in-plane Heisenberg couplings. If we define a particle creation as a spin-flip  on a ferromagnetic background, we can map this model, following standard techniques \cite{Lacroix2011}, into a model of hard-core bosons in an SSH chain with the Hamiltonian of (1), by substituting $-J_i/2 \to t_i$ and $J_z\to V$, where $J_1$, $J_2$ and $J_z$ are, respectively, the two staggered in-plane couplings and the $z$ component coupling.

\subsection{Two-hole states}
\label{subsec:twoparticle}

\begin{figure*}
\begin{center}
\includegraphics[width=0.95 \textwidth ]{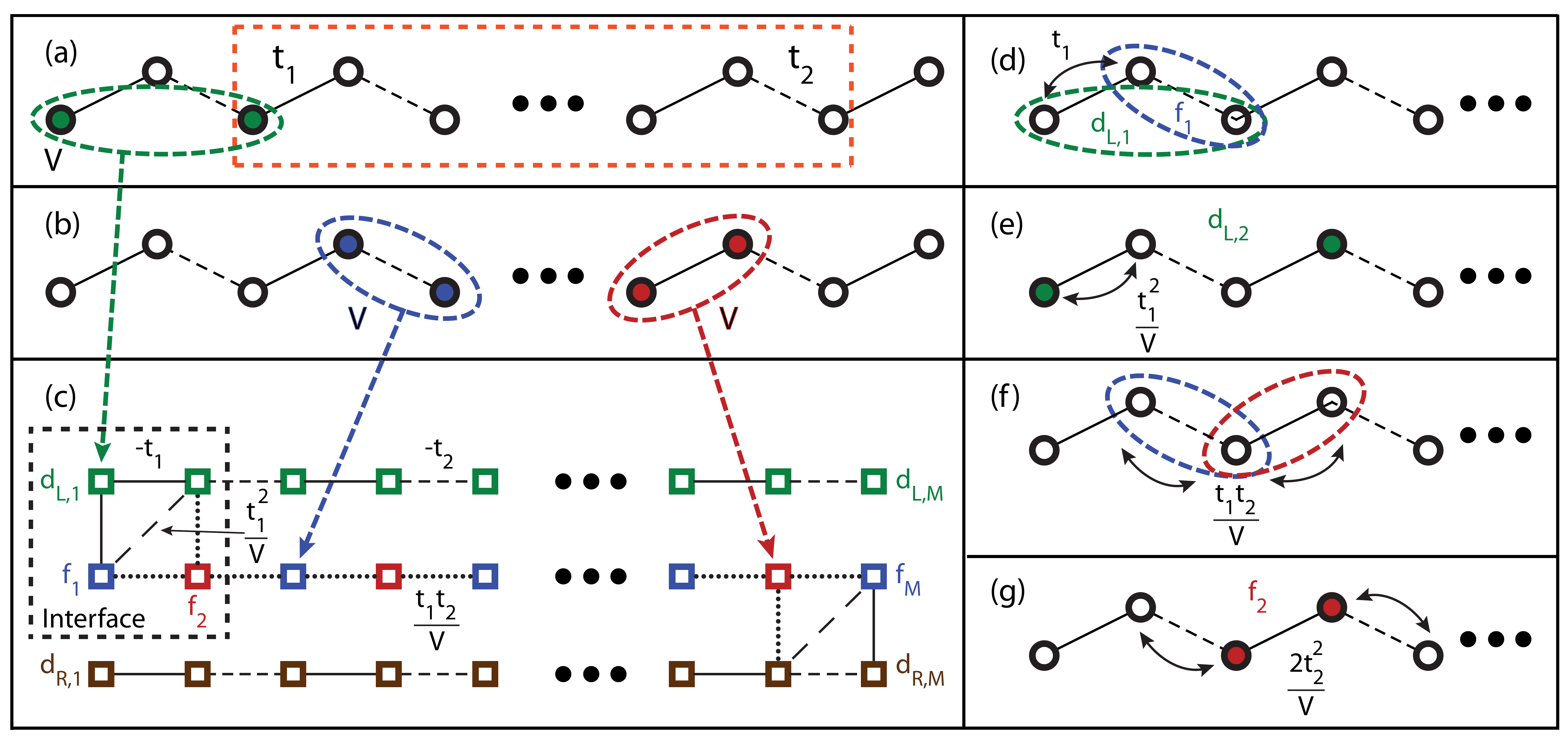}
\end{center}
\caption{(a) and (b): two-hole states with potential energy $V$. 
In (a), one hole is localized at the left edge and the other at any of the sites within the orange dashed box. 
In (b), two examples, with different inner hoppings, of two-hole states with the holes localized at adjacent bulk sites.(c) The two-hole states in (a) and (b) are translated as sites in an equivalent chain, using second-order perturbation theory for $V\gg t_1,t_2$, and can be divided in a (green) $d_L$-chain, a (red and blue) $f$-chain and a (brown) $d_R$ chain.
(d) Hopping constant between $d_{L,1}$ and $f_1$ sites.
(e) On-site potential at $d_{L,2}$.
(f) Hopping constant between nearest neighboring $f$ sites.
(g) On-site potential at $f_2$.
In (d) and (f), occupied sites are not colored due to superposition.}
\label{fig:twoparticles}
\end{figure*}

We have seen that for $\vert V \vert \gg t_1,t_2$, the chain gets effectively shortened for one-hole states [see Fig.~\ref{fig:ssh}(iii)]. 
For two-hole states, we define a new zero of potential, dropping the $V(N-1-4)$ energy constant of the states with two holes localized at non-adjacent bulk sites (where 4 is the number of missing NN interactions in comparison with the half-filled case).
Relative to the new zero of potential, there will be a subspace of states with potential energy $V$ (with 3 missing NN interactions).
In turn, there are two different kinds of states in this subspace. 
One kind has one hole localized at one of the edge sites and the other restricted to the smaller inner SSH chain [orange dashed box in Fig.~\ref{fig:twoparticles}(a)]. 
These states have the form $d_{L(R),i}^{\dagger}\to h_{i+2}^{\dagger}h_{1(N)}^{\dagger}$, $i=1,2...M$ with $M=N-3$, where $d_{L(R),i}^{\dagger}$ has one hole at $h_{1(N)}^\dagger$, the left (right) edge.
The other kind of states consists of two holes localized at adjacent bulk sites, of the form $f_i^{\dagger}\to h_{i+2}^{\dagger}h_{i+1}^{\dagger}$, $i=1,2...M$.
Two different examples are shown in Fig.~\ref{fig:twoparticles}(b).

In the $\vert V \vert \gg t_i$ limit considered, $t_1$ and $t_2$ can be regarded as perturbations that lift degeneracies in the subspace of states with energy $V$. 
By collecting all terms up to second-order, one can construct an equivalent chain, as in Fig.~\ref{fig:twoparticles}(c), where each two-particle state is regarded a site of this chain: $d_{L,i}^{\dagger}$ and $f_i^{\dagger}$ states in Figs.~\ref{fig:twoparticles}(a) and (b) become, respectively, sites of the $d_L$ and $f$ chains in the equivalent chain. 
In what follows, we will only consider the $d_L$ and $f$ chains, which is sufficient to establish the correspondence. 
Apart from small energy corrections at the edge sites, the effect of the introduction of the $d_R$-chain would be to make each energy state doubly degenerate.
Some on-site potential and hopping terms of our second-order expansion are depicted in Figs.~\ref{fig:twoparticles}(d)-(g).
The Hamiltonian of the equivalent chain writes as
\begin{widetext}
\begin{eqnarray}
\label{eq:hamiltonian_equi_chain}
H&=&H_{d_L} + H_f + H_{d_L\leftrightarrow f} + H_V, 
\\
\label{eq:hamiltonian_equi_chain2}
H_{d_L}&=&-t_1 \sum_{j=1}^{\frac{M-1}{2}} \big(d_{L,2j-1}^{\dagger}d_{L,2j} + H.c.\big) - t_2 \sum_{j=1}^{\frac{M-1}{2}}\big(d_{L,2j}^{\dagger}d_{L,2j+1} + H.c.\big) + \frac{t_1^2}{V}\sum_{j=2}^{M-1}n_{j}^{d_L} - \frac{t_1^2}{V} n_1^{d_L} ,
\\
\label{eq:hamiltonian_equi_chain3}
H_f&=&\frac{t_1t_2}{V}\sum_{j=1}^{M-1}\big(f_j^{\dagger}f_{j+1}+H.c.\big) + \frac{2t_1^2}{V}\sum_{j=2}^{\frac{M+1}{2}}n_{2j-1}^f +  \frac{2t_2^2}{V}\sum_{j=1}^{\frac{M+1}{2}}n_{2j}^f + \frac{t_1^2}{V}\big(n_1^f + n_M^f\big),
\\
H_{d_L\leftrightarrow f} &=& -t_1 d_{L,1}^{\dagger}f_1 + \frac{t_1^2}{V} d_{L,2}^{\dagger}f_1 + \frac{t_1t_2}{V}d_{L,2}^{\dagger}f_2 + H.c.
\\
H_V &=& V \sum_{j=1}^M \big(n_j^{d_L}+n_j^f\big), 
\end{eqnarray}
\end{widetext}
where $H_{d_L}$, $H_f$, $H_{d_L\leftrightarrow f}$ and $H_V$ are, respectively, the Hamiltonians of the $d_L$-chain, of the $f$-chain, of their interface and the constant on-site potential attributed at each site, explicitly introduced to keep the correspondence exact.
The potentials at sites $d_{L,1}$ and $d_{L,M}$ are different from the other sites.
For $d_{L,1}$, the intermediate virtual state in the second-order corrections is $h_2^\dagger h_1^\dagger$, with energy $2V$, so the energy difference in the denominator is given by $\Delta E=V-2V=-V$, hence the minus sign in the last term in (\ref{eq:hamiltonian_equi_chain2}). For $d_{L,M}$, two second-order processes are present, one mediated by $h_N^\dagger h_1^\dagger$ with energy $2V$, and the other by $h_{N-1}^\dagger h_2^\dagger$ with energy 0 (meaning $\Delta E=V$), therefore canceling one another.
The complete Hamiltonian of the equivalent chain, with the inclusion of the $d_R$ chain, would have two extra terms written as $H_{d_L}\to H_{d_R}$ and $H_{d_L\leftrightarrow f}\to H_{d_R\leftrightarrow f}$, with the $d_{L,i}(f_i)\to d_{R,M+1-i}(f_{M+1-i})$ substitutions.
$n_1^{d_L}$ and $n_1^f$ are the occupation numbers of the $d_L$ and $f$ chains, respectively.
Because this is the Hamiltonian for one-particle states, and not one-hole states, the hopping terms have reversed signs.
The interface connects two chains with distinct topological natures, since the $f$-chain is trivial and the $d_L$-chain non-trivial (the usual SSH chain with on-site potentials and mixed edges).

We studied a concrete example of two-hole states in an SSH chain with $N=20$ sites and compared them to the states present in the equivalent chain with 34 sites (both the $d_L$-chain and the $f$-chain have $M=N-3=17$ sites).
The states of the SSH and equivalent chain are written, respectively, as
\begin{eqnarray}
\vert \psi_{ SSH} \rangle &=& \sum_{i=1}^{N-1} \sum_{j=i+1}^N\alpha_{ij} h_j^{\dagger}h_i^{\dagger}\vert \emptyset_{ssh} \rangle,
\\ 
&&\sum_{i=1}^{N-1} \sum_{j=i+1}^N\vert\alpha_{ij}\vert^2=1, \nonumber
\\
\vert \psi_{chain} \rangle &=& \sum_{j=1}^M \big(\beta_j d_{L,j}^{\dagger} + \gamma_j f_j^{\dagger} \big)\vert \emptyset_{chain} \rangle,
\\
&&\sum_{j=1}^M \big(\vert\beta_j\vert^2 + \vert\gamma_j\vert^2 \big)=1. \nonumber
\end{eqnarray}
After finding the coefficients numerically, the mean occupation for each site is given by $\langle n_j\rangle = \langle \psi \vert h_j^{\dagger}h_j \vert \psi \rangle$. 
In the case of the equivalent chain, we use the correspondences $d_{L,i}^{\dagger}\vert \emptyset_{chain} \rangle \to h_{i+2}^{\dagger}h_{1}^{\dagger}\vert \emptyset_{ssh} \rangle$ and $f_i^{\dagger}\vert \emptyset_{chain} \rangle \to h_{i+2}^{\dagger}h_{i+1}^{\dagger}\vert \emptyset_{ssh} \rangle$.
Note that $\langle n\rangle = \sum_{j=1}^N \langle n_j\rangle = 2$.

\begin{figure*}[t]
\begin{center}
\includegraphics[width=0.98 \textwidth ,height=0.35 \textheight]{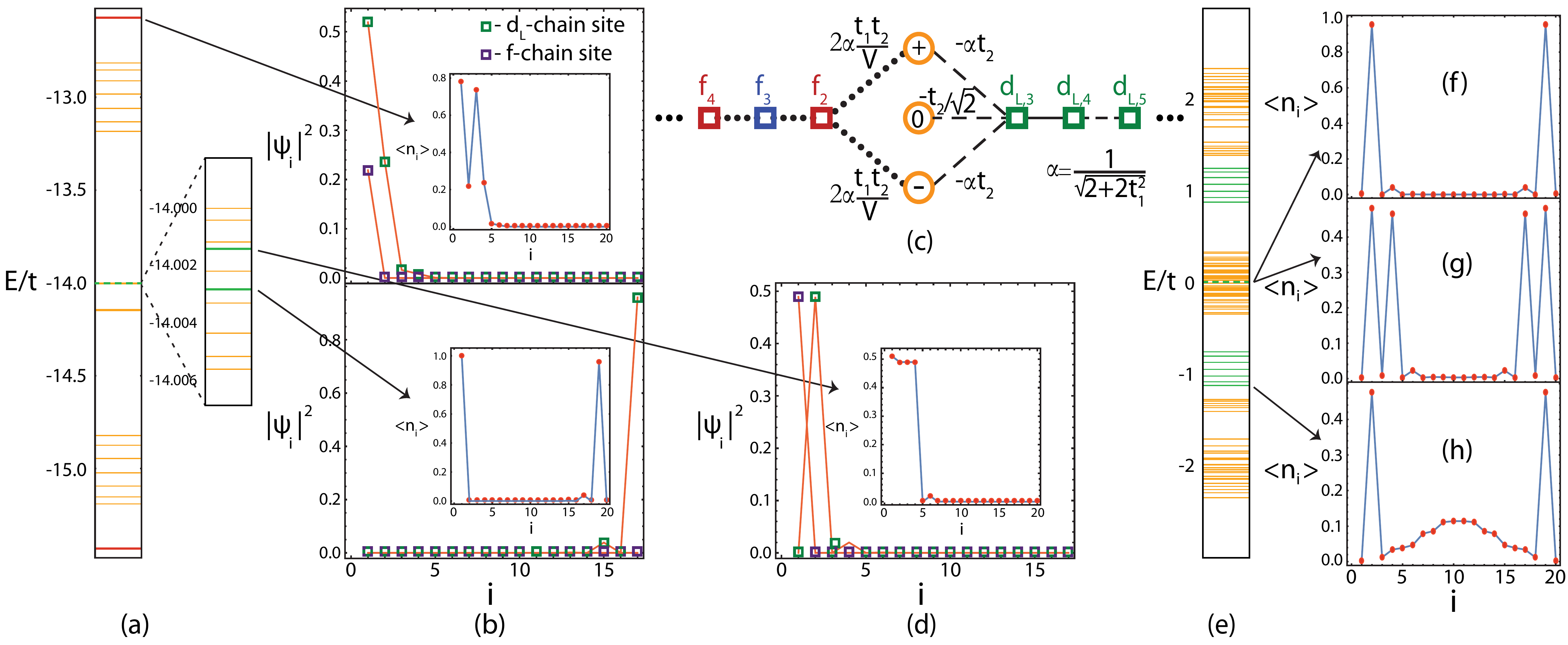}
\end{center}
\caption{(a) Energy spectrum for two-hole states in a chain with $N=20$ sites, in the subspace of states with potential energy $V$. 
Parameters are $(V,t_1,t_2) \to t(-14,1,0.2)$.
Each energy level is doubly degenerate. 
Red, dashed green and orange levels represent, respectively, impurity-like, topologically originated and itinerant bulk states. 
There are two (doubly degenerate) topological states localized around $E/t=-14$, superposed with a very narrow band of bulk states, shown in the zoomed region.
(b) Probability amplitude squared at sites $i$ of the equivalent chain for the impurity-like (top) and one of the topological (bottom) states indicated by the arrows.
The insets show how these states of the equivalent chain translate in terms of mean occupation in the original SSH chain.
Their corresponding degenerate states are given by a reflection about the center of the chain.
(c) For $t_2\ll t_1\ll V$, the diagonalization of the Hamiltonian written in the $\{\ket{f_1},\ket{d_{L,1}},\ket{d_{L,2}}\}$ basis of the equivalent chain in Fig.~\ref{fig:twoparticles}(c) yields three different states at the interface, $\ket{0}=\frac{1}{\sqrt{2}}(-1,0,1)$ and $\ket{\pm}\approx\frac{1}{\sqrt{2+2t_1^2}}(1,\pm \sqrt{2 t_1},1)$, with energies $E_0=0$ and $E_{\pm}\approx \frac{t_1^2}{2V}\pm \sqrt{2}t_1$ (the on-site potentials at these new interface ``sites''), considering $t_1\approx 1$. The hoppings form these new ``sites'' to the $f$ and $d_L$ chains become renormalized. 
(d) Same as in (b), with $(V,t_1,t_2) \to t(-14,1,0.2)$, for the state of topological origin with a large weight on $\ket{0}$, with zero energy [or at $E/t=-14$ in Fig.~\ref{fig:energylevelstwoparticles}(a) when the constant potential $V$ is considered].
(e) Same as in (a) but for the subspace of zero potential energy. All states in the green bands have one of the holes localized at the first inner site, as in (h).
(f)-(h) Mean occupation of the two-hole states of topological origin indicated by the arrows. 
(f) One hole localized at the first inner site of each end. 
(g) One hole localized at the first and another at the third inner site. 
(f) and (g) are  not topologically protected, as they are buried in a continuum of other itinerant states.
(h) One hole localized at the first inner site, topologically protected, and another in an itinerant bulk state. 
States (g) and (h) are given by a linear combination of two degenerate eigenstates.
In all localized holes a decaying tail to the bulk is implied.}
\label{fig:energylevelstwoparticles}
\end{figure*}

For $(V,t_1,t_2) \to t(-14,1,0.2)$, where $t$ is an arbitrary energy constant, the energy spectrum for the SSH chain with $N=20$ sites, in the subspace of states with potential energy $V$ is shown in Fig.~\ref{fig:energylevelstwoparticles}(a).
Bands around $E/t=-13$ and $E/t=-15$ correspond to itinerant bulk states of the hole confined to the chain delimited by the orange dashed box in Fig.~\ref{fig:twoparticles}(a), that is, bulk states in the corresponding $d$ chains, while
very narrow bands around $E/t=-14$ and $E/t\approx -14.14$ correspond to itinerant bulk states for the bound states of Fig.~\ref{fig:twoparticles}(b), that is, bulk states in the corresponding $f$ chain.
The appearance of a gap between these narrow bands, $\Delta E\approx 0.14t \approx \frac{2t_1^2}{V}$, is explained by the alternating on-site potentials in the $f$ chain [see (\ref{eq:hamiltonian_equi_chain3})].
Two impurity-like states (red levels with $E/t\simeq -15.49$ and $E/t\simeq -12.65$), whose origin will be discussed below, and two states of topological origin are found (dashed green levels at $E/t=-14$, superposed with other non-topological orange levels consisting of bulk states in the $f$ chain). 
In what follows, we need to differentiate states with topological origin, but not topologically protected, from true topological states that are robust against disorder, \textit{i.e.}, that cannot couple and scatter into bulk states. Regardless of this distinction, the topological origin of all these states stems from the underlying SSH geometry present in the equivalent chain.

The set of parameters considered is enough to ensure that one is, for all intended purposes, in the $V\gg t_1,t_2$ limit.
Therefore, one can construct the equivalent chain, using the Hamiltonian in (\ref{eq:hamiltonian_equi_chain}), and verify the correspondence between its one-particle states and the two-hole states in the original SSH chain.
The energy spectrum of one-particle states in the equivalent chain was found to match, apart from negligible energy corrections, that of Fig.~\ref{fig:energylevelstwoparticles}(a).
The profiles of the top impurity-like state (red level with $E/t\simeq -12.65$) and of one of the edge states with a topological origin (dashed green level with $E/t=-14$), in the equivalent chain, are shown in Fig.~\ref{fig:energylevelstwoparticles}(b).
The insets show how they translate in terms of mean occupancy at the sites of the original SSH chain. 
The state shown in the bottom inset has one hole localized at the left edge and the other in a topological state with a decaying tail from the right edge of the smaller inner chain [see orange dashed box in Fig.~\ref{fig:twoparticles}(a)], which ends with a $t_2$ hopping.
But is this state topologically protected, given that its energy, at $E/t=-14$ [see Fig.~\ref{fig:energylevelstwoparticles}(a)], is inside the narrow band of bulk states of the $f$ chain (superposed orange levels)?
Even though there is no gap separating it from the $f$ bulk states, one can still argue that it can be regarded as a topologically protected state. 
This can be better understood by looking at the equivalent chain in Fig.~\ref{fig:twoparticles}(c). 
The topological state is localized at the right edge of the $d_L$ chain, while the others are bulk states of the $f$ chain. 
To perturbatively couple $d_{L,M}$ with any of the $f$ sites would require long-range hoppings which, for a $d_L$ chain larger then only a few sites, becomes physically unrealistic.
The same reasoning applies to the introduction of local disorder, against which this state is also protected.
The profiles of each inset reveal an almost exact agreement with the states of the original chain, which confirms that the problem of one-particle states in the equivalent chain captures the essential features of the two-hole problem in the SSH chain, in the limit considered.

We mentioned the existence of two states of topological origin for $(V,t_1,t_2) \to t(-14,1,0.2)$. 
Their energies are quasi-degenerate and so they are both represented by the green level in Fig.~\ref{fig:energylevelstwoparticles}(a) at $E/t=-14$.
One of them, shown at the bottom inset of Fig.~\ref{fig:energylevelstwoparticles}(b), has a hole located at the right edge of the orange dashed box in Fig.~\ref{fig:twoparticles}(a), which ends in a $t_2$ hopping, as expected.
Since the left edge ends with a $t_1$ hopping, how can a second state of topological origin be present at this edge, given that $t_2<t_1$?
Suppose $t_2\ll t_1 \ll V$. 
In this limit, terms with $t_2$ can be treated as a perturbation in the equivalent chain in Fig.~\ref{fig:twoparticles}(c). 
Let us focus on the three sites at the interface, which form the basis $\{\ket{f_1},\ket{d_{L,1}},\ket{d_{L,2}}\}$, connected to the rest of the chain by terms containing $t_2$ [see dashed box in Fig.~\ref{fig:twoparticles}(c)], and diagonalize them first. The Hamiltonian in this basis is given by
\begin{equation}
H=
\begin{pmatrix}
t_1^2/V & t_1 & t_1^2/V
\\
t_1 & -t_1^2/V & t_1
\\
t_1^2/V & t_1 & t_1^2/V
\end{pmatrix},
\end{equation}
where a constant on-site potential term, $V \hat{I}$, with $\hat{I}$ the identity matrix, was taken out.
Diagonalization yields the following eigenvalues and eigenvectors: $E_{\pm}\approx \frac{t_1^2}{2V}\pm \sqrt{2}t_1$ and $E_0=0$, with $\ket{\pm}\approx\frac{1}{\sqrt{2+2t_1^2}}(1,\pm \sqrt{2 t_1},1)$, considering $t_1\approx 1$, and $\ket{0}=\frac{1}{\sqrt{2}}(-1,0,1)$.
When $t_2$ is turned on, these are the three hopping possibilities available at the interface, as shown in Fig.~\ref{fig:energylevelstwoparticles}(c). 
Notice that $\ket{0}$ has no component in $\ket{d_{L,1}}$, so that for this case the $d_L$-chain effectively starts at the $d_{L,2}$ site, that is, the hopping at the left edge of this shorter $d_L$-chain, now between $\ket{0}$ and $\ket{d_{L,3}}$, is given by $\frac{t_2}{\sqrt{2}}$, which allows for the presence of a second state of topological origin, with the profile of Fig.~\ref{fig:energylevelstwoparticles}(d), which is not protected since it couples to the $f$ chain \textit{via} $\ket{f_2}$ when perturbations are introduced.
Furthermore, ``sites'' $\ket{\pm}$ can be readily identified as the origin of the impurity-like states lying at the interface [red levels in Fig.~\ref{fig:energylevelstwoparticles}(a) with energies $E/t\simeq -12.65$ and $E/t\simeq -15.49$, when the constant $V$ is added].
On-site potentials $E_\pm>|t_2|$ at $\ket{\pm}$ throw these two new left ends of the $d_L$ chain into the non-topological sector in the phase diagram of Fig.~\ref{fig:ssh}(iv).
Even though $(V,t_1,t_2) \to t(-14,1,0.2)$ is still somewhat far from the $t_2\ll t_1 \ll V$ limit, the above considerations are enough to account, nonetheless, for the results obtained.

The states available in the zero potential subspace, which yield the energy spectrum of Fig.~\ref{fig:energylevelstwoparticles}(e), are composed of two non-consecutive bulk holes.
Thus, as in Fig.~\ref{fig:ssh}(iii), the two holes are confined to the inner chain with $t_2$ hoppings at its ends.
Under these conditions, three additional kinds of states of topological origin can be found in this subspace: one with one hole localized at each end of the inner chain [see Fig.~\ref{fig:energylevelstwoparticles}(f)], another with the two holes localized at the edge and at the first non-consecutive site of the inner chain [see Fig.~\ref{fig:energylevelstwoparticles}(g)], and finally one with one hole localized at an edge of the inner chain and one hole in an itinerant state at the bulk [see Fig.~\ref{fig:energylevelstwoparticles}(h)].
Only this last one can be considered topologically protected, in the following sense: 
while the states in Fig.~\ref{fig:energylevelstwoparticles}(f) and (g) are buried in a continuum of itinerant states around zero energy, the states of the type of Fig.~\ref{fig:energylevelstwoparticles}(h) form two bands (the itinerant hole at the bulk gives the usual SSH energy spectrum, since the hole at the edge has zero energy) energetically separated from the others, as shown in Fig.~\ref{fig:energylevelstwoparticles}(e). 
When disorder is introduced in a state within one of these two bands, its effect will be to scatter the bulk hole, while the edge hole remains unaffected, since it is a common feature of all the states in that band.
Therefore, the edge hole of Fig.~\ref{fig:energylevelstwoparticles}(h) is topologically protected.
However, note that we may think of other definitions of topological protection of many-body states.
In particular, a more severe definition would be to require that the complete many-body state is protected against disorder \cite{Grusdt2013}.
In this paper, we adopt the less severe definition explained above.
Note that throughout this paper we consider as topologically originated all multi-hole states where at least one hole is in a state with topological origin.

When the values of the hoppings are switched, $(V,t_1,t_2) \to t(-14,0.2,1)$, one finds two different states of topological origin, both unprotected, and no impurity-like states.
These states have one of the holes localized at the left edge of the orange dashed box in Fig.~\ref{fig:twoparticles}(a), since  $t_1<t_2$.
Starting with $t_1=0$, the $d_{L,1}$ and $f_1$ sites are isolated, and both states of the equivalent chain with one particle located in one of these sites have energy $V$.
Using the same reasoning as before, one takes now the $t_1\ll t_2 \ll V$ limit, where $t_1$ is a perturbation. 
As $t_1$ is turned on, the degeneracy of these states is lifted first, yielding new states that are a symmetric and anti-symmetric combination of \{$\ket{d_{L,1}},\ket{f_1}$\}.
Both these new states have a component in $\ket{d_{L,1}}$, which is linked to $\ket{d_{L,2}}$ by $\frac{t_1}{\sqrt{2}}>t_2$, that is, one effectively has two left edges that support two states of topological origin, although unprotected due to the $\ket{f_1}$ component.

Our results show that, for high enough $V$, two-hole states of topological origin are present in the SSH chain in both the $\frac{t_1}{t_2}\lessgtr 1$ regimes. This comes as a consequence of the mixed edges of the smaller chain [see orange dashed box in Fig.~\ref{fig:twoparticles}(a)] to which one of the holes is restricted.

\section{Conclusions}
\label{sec:conclusion}

The natural extension of our results would be the characterization of topologically originated $n$-hole states, with $n>2$. 
As more holes are introduced into the chain, more energy subspaces are available and one expects the appearance of a cascade of topologically originated states distributed in these subspaces, until $n=N/2$. 
From this point onwards, the introduction of more holes decreases the energy subspaces available and, therefore, the number of topologically originated states present. 
Note that, for $n=N-1$ holes in the chain, one should recover the classical one-particle energy spectrum of the SSH chain.
For $V\gg t_1>t_2$ and $n\leq N/2-1$, there should be a particular topologically originated state at the zero energy subspace with $n$ holes localized at every other site starting from the first inner site, as shown in Fig.~\ref{fig:energylevelstwoparticles}(g) for $n=2$.
A similar state should also be present for $V\gg t_2>t_1$ and $n\leq N/2$ in the $V$ energy subspace, where one of the non-consecutive holes is at an edge site.
One also expects the correspondence between $n$-hole topological states and one-particle states in equivalent chains to hold, whose construction rules are specific to each case, following the results drawn here for two-hole states [see Fig.~\ref{fig:twoparticles}(c)].



\section*{Acknowledgments}\label{sec:acknowledments}

This work is funded by FEDER funds through the COMPETE 2020 Programme and National Funds throught FCT - Portuguese Foundation for Science and Technology under the project UID/CTM/50025/2013.
AMM acknowledges the financial support from the FCT through the grant SFRH/PD/BD/108663/2015.
RGD thanks the support by the Beijing CSRC.


\bibliography{topologicalinteraction}

\begin{thebibliography}{38}%
\makeatletter
\providecommand \@ifxundefined [1]{%
 \@ifx{#1\undefined}
}%
\providecommand \@ifnum [1]{%
 \ifnum #1\expandafter \@firstoftwo
 \else \expandafter \@secondoftwo
 \fi
}%
\providecommand \@ifx [1]{%
 \ifx #1\expandafter \@firstoftwo
 \else \expandafter \@secondoftwo
 \fi
}%
\providecommand \natexlab [1]{#1}%
\providecommand \enquote  [1]{``#1''}%
\providecommand \bibnamefont  [1]{#1}%
\providecommand \bibfnamefont [1]{#1}%
\providecommand \citenamefont [1]{#1}%
\providecommand \href@noop [0]{\@secondoftwo}%
\providecommand \href [0]{\begingroup \@sanitize@url \@href}%
\providecommand \@href[1]{\@@startlink{#1}\@@href}%
\providecommand \@@href[1]{\endgroup#1\@@endlink}%
\providecommand \@sanitize@url [0]{\catcode `\\12\catcode `\$12\catcode
  `\&12\catcode `\#12\catcode `\^12\catcode `\_12\catcode `\%12\relax}%
\providecommand \@@startlink[1]{}%
\providecommand \@@endlink[0]{}%
\providecommand \url  [0]{\begingroup\@sanitize@url \@url }%
\providecommand \@url [1]{\endgroup\@href {#1}{\urlprefix }}%
\providecommand \urlprefix  [0]{URL }%
\providecommand \Eprint [0]{\href }%
\providecommand \doibase [0]{http://dx.doi.org/}%
\providecommand \selectlanguage [0]{\@gobble}%
\providecommand \bibinfo  [0]{\@secondoftwo}%
\providecommand \bibfield  [0]{\@secondoftwo}%
\providecommand \translation [1]{[#1]}%
\providecommand \BibitemOpen [0]{}%
\providecommand \bibitemStop [0]{}%
\providecommand \bibitemNoStop [0]{.\EOS\space}%
\providecommand \EOS [0]{\spacefactor3000\relax}%
\providecommand \BibitemShut  [1]{\csname bibitem#1\endcsname}%
\let\auto@bib@innerbib\@empty
\bibitem [{\citenamefont {Su}\ \emph {et~al.}(1979)\citenamefont {Su},
  \citenamefont {Schrieffer},\ and\ \citenamefont {Heeger}}]{Su1979}%
  \BibitemOpen
  \bibfield  {author} {\bibinfo {author} {\bibfnamefont {W.~P.}\ \bibnamefont
  {Su}}, \bibinfo {author} {\bibfnamefont {J.~R.}\ \bibnamefont {Schrieffer}},
  \ and\ \bibinfo {author} {\bibfnamefont {A.~J.}\ \bibnamefont {Heeger}},\
  }\href {\doibase 10.1103/PhysRevLett.42.1698} {\bibfield  {journal} {\bibinfo
   {journal} {Phys. Rev. Lett.}\ }\textbf {\bibinfo {volume} {42}},\ \bibinfo
  {pages} {1698} (\bibinfo {year} {1979})}\BibitemShut {NoStop}%
\bibitem [{\citenamefont {Asb\'oth}\ \emph {et~al.}(2016)\citenamefont
  {Asb\'oth}, \citenamefont {Oroszl\'any},\ and\ \citenamefont
  {P\'alyi}}]{Asboth2016}%
  \BibitemOpen
  \bibfield  {author} {\bibinfo {author} {\bibfnamefont {K.~J.}\ \bibnamefont
  {Asb\'oth}}, \bibinfo {author} {\bibfnamefont {L.}~\bibnamefont
  {Oroszl\'any}}, \ and\ \bibinfo {author} {\bibfnamefont {A.}~\bibnamefont
  {P\'alyi}},\ }\href@noop {} {\emph {\bibinfo {title} {A Short Course on
  Topological Insulators}}},\ edited by\ \bibinfo {editor} {\bibnamefont
  {Springer}}\ (\bibinfo  {publisher} {Lecture Notes in Physics},\ \bibinfo
  {year} {2016})\BibitemShut {NoStop}%
\bibitem [{\citenamefont {Guo}\ and\ \citenamefont {Shen}(2011)}]{Guo2011}%
  \BibitemOpen
  \bibfield  {author} {\bibinfo {author} {\bibfnamefont {H.}~\bibnamefont
  {Guo}}\ and\ \bibinfo {author} {\bibfnamefont {S.-Q.}\ \bibnamefont {Shen}},\
  }\href {\doibase 10.1103/PhysRevB.84.195107} {\bibfield  {journal} {\bibinfo
  {journal} {Phys. Rev. B}\ }\textbf {\bibinfo {volume} {84}},\ \bibinfo
  {pages} {195107} (\bibinfo {year} {2011})}\BibitemShut {NoStop}%
\bibitem [{\citenamefont {Niu}\ and\ \citenamefont {Thouless}(1984)}]{Niu1984}%
  \BibitemOpen
  \bibfield  {author} {\bibinfo {author} {\bibfnamefont {Q.}~\bibnamefont
  {Niu}}\ and\ \bibinfo {author} {\bibfnamefont {D.~J.}\ \bibnamefont
  {Thouless}},\ }\href {http://stacks.iop.org/0305-4470/17/i=12/a=016}
  {\bibfield  {journal} {\bibinfo  {journal} {J. Phys. A: Math. Gen.}\ }\textbf
  {\bibinfo {volume} {17}},\ \bibinfo {pages} {2453} (\bibinfo {year}
  {1984})}\BibitemShut {NoStop}%
\bibitem [{\citenamefont {Xiao}\ \emph {et~al.}(2010)\citenamefont {Xiao},
  \citenamefont {Chang},\ and\ \citenamefont {Niu}}]{Xiao2010}%
  \BibitemOpen
  \bibfield  {author} {\bibinfo {author} {\bibfnamefont {D.}~\bibnamefont
  {Xiao}}, \bibinfo {author} {\bibfnamefont {M.-C.}\ \bibnamefont {Chang}}, \
  and\ \bibinfo {author} {\bibfnamefont {Q.}~\bibnamefont {Niu}},\ }\href
  {\doibase 10.1103/RevModPhys.82.1959} {\bibfield  {journal} {\bibinfo
  {journal} {Rev. Mod. Phys.}\ }\textbf {\bibinfo {volume} {82}},\ \bibinfo
  {pages} {1959} (\bibinfo {year} {2010})}\BibitemShut {NoStop}%
\bibitem [{\citenamefont {Guo}\ \emph {et~al.}(2012)\citenamefont {Guo},
  \citenamefont {Shen},\ and\ \citenamefont {Feng}}]{Guo2012}%
  \BibitemOpen
  \bibfield  {author} {\bibinfo {author} {\bibfnamefont {H.}~\bibnamefont
  {Guo}}, \bibinfo {author} {\bibfnamefont {S.-Q.}\ \bibnamefont {Shen}}, \
  and\ \bibinfo {author} {\bibfnamefont {S.}~\bibnamefont {Feng}},\ }\href
  {\doibase 10.1103/PhysRevB.86.085124} {\bibfield  {journal} {\bibinfo
  {journal} {Phys. Rev. B}\ }\textbf {\bibinfo {volume} {86}},\ \bibinfo
  {pages} {085124} (\bibinfo {year} {2012})}\BibitemShut {NoStop}%
\bibitem [{\citenamefont {Budich}\ and\ \citenamefont
  {Ardonne}(2013)}]{Budich2013}%
  \BibitemOpen
  \bibfield  {author} {\bibinfo {author} {\bibfnamefont {J.~C.}\ \bibnamefont
  {Budich}}\ and\ \bibinfo {author} {\bibfnamefont {E.}~\bibnamefont
  {Ardonne}},\ }\href {\doibase 10.1103/PhysRevB.88.035139} {\bibfield
  {journal} {\bibinfo  {journal} {Phys. Rev. B}\ }\textbf {\bibinfo {volume}
  {88}},\ \bibinfo {pages} {035139} (\bibinfo {year} {2013})}\BibitemShut
  {NoStop}%
\bibitem [{\citenamefont {Kivelson}\ and\ \citenamefont
  {Heim}(1982)}]{Kivelson1982}%
  \BibitemOpen
  \bibfield  {author} {\bibinfo {author} {\bibfnamefont {S.}~\bibnamefont
  {Kivelson}}\ and\ \bibinfo {author} {\bibfnamefont {D.~E.}\ \bibnamefont
  {Heim}},\ }\href {\doibase 10.1103/PhysRevB.26.4278} {\bibfield  {journal}
  {\bibinfo  {journal} {Phys. Rev. B}\ }\textbf {\bibinfo {volume} {26}},\
  \bibinfo {pages} {4278} (\bibinfo {year} {1982})}\BibitemShut {NoStop}%
\bibitem [{\citenamefont {Glick}\ and\ \citenamefont
  {Bryant}(1986)}]{Glick1986}%
  \BibitemOpen
  \bibfield  {author} {\bibinfo {author} {\bibfnamefont {A.~J.}\ \bibnamefont
  {Glick}}\ and\ \bibinfo {author} {\bibfnamefont {G.~W.}\ \bibnamefont
  {Bryant}},\ }\href {\doibase 10.1103/PhysRevB.34.943} {\bibfield  {journal}
  {\bibinfo  {journal} {Phys. Rev. B}\ }\textbf {\bibinfo {volume} {34}},\
  \bibinfo {pages} {943} (\bibinfo {year} {1986})}\BibitemShut {NoStop}%
\bibitem [{\citenamefont {Glick}\ \emph {et~al.}(1988)\citenamefont {Glick},
  \citenamefont {Cohen},\ and\ \citenamefont {Bryant}}]{Glick1988}%
  \BibitemOpen
  \bibfield  {author} {\bibinfo {author} {\bibfnamefont {A.~J.}\ \bibnamefont
  {Glick}}, \bibinfo {author} {\bibfnamefont {R.~J.}\ \bibnamefont {Cohen}}, \
  and\ \bibinfo {author} {\bibfnamefont {G.~W.}\ \bibnamefont {Bryant}},\
  }\href {\doibase 10.1103/PhysRevB.37.2653} {\bibfield  {journal} {\bibinfo
  {journal} {Phys. Rev. B}\ }\textbf {\bibinfo {volume} {37}},\ \bibinfo
  {pages} {2653} (\bibinfo {year} {1988})}\BibitemShut {NoStop}%
\bibitem [{\citenamefont {Rossi}(1992)}]{Rossi1992}%
  \BibitemOpen
  \bibfield  {author} {\bibinfo {author} {\bibfnamefont {G.}~\bibnamefont
  {Rossi}},\ }\href {\doibase http://dx.doi.org/10.1016/0379-6779(92)90093-X}
  {\bibfield  {journal} {\bibinfo  {journal} {Synt. Met.}\ }\textbf {\bibinfo
  {volume} {49}},\ \bibinfo {pages} {221 } (\bibinfo {year}
  {1992})}\BibitemShut {NoStop}%
\bibitem [{\citenamefont {Yan}\ and\ \citenamefont {Wan}(2014)}]{Yan2014}%
  \BibitemOpen
  \bibfield  {author} {\bibinfo {author} {\bibfnamefont {Z.}~\bibnamefont
  {Yan}}\ and\ \bibinfo {author} {\bibfnamefont {S.}~\bibnamefont {Wan}},\
  }\href {http://stacks.iop.org/0295-5075/107/i=4/a=47007} {\bibfield
  {journal} {\bibinfo  {journal} {EPL}\ }\textbf {\bibinfo {volume} {107}},\
  \bibinfo {pages} {47007} (\bibinfo {year} {2014})}\BibitemShut {NoStop}%
\bibitem [{\citenamefont {Sticlet}\ \emph {et~al.}(2014)\citenamefont
  {Sticlet}, \citenamefont {Seabra}, \citenamefont {Pollmann},\ and\
  \citenamefont {Cayssol}}]{Sticlet2014}%
  \BibitemOpen
  \bibfield  {author} {\bibinfo {author} {\bibfnamefont {D.}~\bibnamefont
  {Sticlet}}, \bibinfo {author} {\bibfnamefont {L.}~\bibnamefont {Seabra}},
  \bibinfo {author} {\bibfnamefont {F.}~\bibnamefont {Pollmann}}, \ and\
  \bibinfo {author} {\bibfnamefont {J.}~\bibnamefont {Cayssol}},\ }\href
  {\doibase 10.1103/PhysRevB.89.115430} {\bibfield  {journal} {\bibinfo
  {journal} {Phys. Rev. B}\ }\textbf {\bibinfo {volume} {89}},\ \bibinfo
  {pages} {115430} (\bibinfo {year} {2014})}\BibitemShut {NoStop}%
\bibitem [{\citenamefont {Weber}\ \emph {et~al.}(2015)\citenamefont {Weber},
  \citenamefont {Assaad},\ and\ \citenamefont {Hohenadler}}]{Weber2015}%
  \BibitemOpen
  \bibfield  {author} {\bibinfo {author} {\bibfnamefont {M.}~\bibnamefont
  {Weber}}, \bibinfo {author} {\bibfnamefont {F.~F.}\ \bibnamefont {Assaad}}, \
  and\ \bibinfo {author} {\bibfnamefont {M.}~\bibnamefont {Hohenadler}},\
  }\href {\doibase 10.1103/PhysRevB.91.245147} {\bibfield  {journal} {\bibinfo
  {journal} {Phys. Rev. B}\ }\textbf {\bibinfo {volume} {91}},\ \bibinfo
  {pages} {245147} (\bibinfo {year} {2015})}\BibitemShut {NoStop}%
\bibitem [{\citenamefont {Ma}\ and\ \citenamefont
  {Schollwöck}(2009)}]{Ma2009}%
  \BibitemOpen
  \bibfield  {author} {\bibinfo {author} {\bibfnamefont {H.}~\bibnamefont
  {Ma}}\ and\ \bibinfo {author} {\bibfnamefont {U.}~\bibnamefont
  {Schollwöck}},\ }\href {\doibase 10.1021/jp809045r} {\bibfield  {journal}
  {\bibinfo  {journal} {J. Phys. Chem. A}\ }\textbf {\bibinfo {volume} {113}},\
  \bibinfo {pages} {1360} (\bibinfo {year} {2009})}\BibitemShut {NoStop}%
\bibitem [{\citenamefont {da~Cunha}\ \emph {et~al.}(2014)\citenamefont
  {da~Cunha}, \citenamefont {Ribeiro~Junior}, \citenamefont {Gargano},\ and\
  \citenamefont {e~Silva}}]{Cunha2014}%
  \BibitemOpen
  \bibfield  {author} {\bibinfo {author} {\bibfnamefont {W.~F.}\ \bibnamefont
  {da~Cunha}}, \bibinfo {author} {\bibfnamefont {L.~A.}\ \bibnamefont
  {Ribeiro~Junior}}, \bibinfo {author} {\bibfnamefont {R.}~\bibnamefont
  {Gargano}}, \ and\ \bibinfo {author} {\bibfnamefont {G.~M.}\ \bibnamefont
  {e~Silva}},\ }\href {\doibase 10.1039/C4CP02184C} {\bibfield  {journal}
  {\bibinfo  {journal} {Phys. Chem. Chem. Phys.}\ }\textbf {\bibinfo {volume}
  {16}},\ \bibinfo {pages} {17072} (\bibinfo {year} {2014})}\BibitemShut
  {NoStop}%
\bibitem [{\citenamefont {Zhang}\ \emph {et~al.}(2016)\citenamefont {Zhang},
  \citenamefont {Liu},\ and\ \citenamefont {An}}]{Zhang2016}%
  \BibitemOpen
  \bibfield  {author} {\bibinfo {author} {\bibfnamefont {Y.}~\bibnamefont
  {Zhang}}, \bibinfo {author} {\bibfnamefont {X.}~\bibnamefont {Liu}}, \ and\
  \bibinfo {author} {\bibfnamefont {Z.}~\bibnamefont {An}},\ }\href {\doibase
  http://dx.doi.org/10.1016/j.orgel.2015.10.010} {\bibfield  {journal}
  {\bibinfo  {journal} {Org Electron}\ }\textbf {\bibinfo {volume} {28}},\
  \bibinfo {pages} {6 } (\bibinfo {year} {2016})}\BibitemShut {NoStop}%
\bibitem [{\citenamefont {Grusdt}\ \emph {et~al.}(2013)\citenamefont {Grusdt},
  \citenamefont {H\"oning},\ and\ \citenamefont {Fleischhauer}}]{Grusdt2013}%
  \BibitemOpen
  \bibfield  {author} {\bibinfo {author} {\bibfnamefont {F.}~\bibnamefont
  {Grusdt}}, \bibinfo {author} {\bibfnamefont {M.}~\bibnamefont {H\"oning}}, \
  and\ \bibinfo {author} {\bibfnamefont {M.}~\bibnamefont {Fleischhauer}},\
  }\href {\doibase 10.1103/PhysRevLett.110.260405} {\bibfield  {journal}
  {\bibinfo  {journal} {Phys. Rev. Lett.}\ }\textbf {\bibinfo {volume} {110}},\
  \bibinfo {pages} {260405} (\bibinfo {year} {2013})}\BibitemShut {NoStop}%
\bibitem [{\citenamefont {Liu}\ \emph {et~al.}(2013)\citenamefont {Liu},
  \citenamefont {Liu},\ and\ \citenamefont {Cheng}}]{Liu2013}%
  \BibitemOpen
  \bibfield  {author} {\bibinfo {author} {\bibfnamefont {X.-J.}\ \bibnamefont
  {Liu}}, \bibinfo {author} {\bibfnamefont {Z.-X.}\ \bibnamefont {Liu}}, \ and\
  \bibinfo {author} {\bibfnamefont {M.}~\bibnamefont {Cheng}},\ }\href
  {\doibase 10.1103/PhysRevLett.110.076401} {\bibfield  {journal} {\bibinfo
  {journal} {Phys. Rev. Lett.}\ }\textbf {\bibinfo {volume} {110}},\ \bibinfo
  {pages} {076401} (\bibinfo {year} {2013})}\BibitemShut {NoStop}%
\bibitem [{\citenamefont {Li}\ and\ \citenamefont {Chen}(2015)}]{Li2015}%
  \BibitemOpen
  \bibfield  {author} {\bibinfo {author} {\bibfnamefont {L.}~\bibnamefont
  {Li}}\ and\ \bibinfo {author} {\bibfnamefont {S.}~\bibnamefont {Chen}},\
  }\href {http://stacks.iop.org/0295-5075/109/i=4/a=40006} {\bibfield
  {journal} {\bibinfo  {journal} {EPL}\ }\textbf {\bibinfo {volume} {109}},\
  \bibinfo {pages} {40006} (\bibinfo {year} {2015})}\BibitemShut {NoStop}%
\bibitem [{\citenamefont {Valiente}\ and\ \citenamefont
  {Petrosyan}(2008)}]{Valiente2008}%
  \BibitemOpen
  \bibfield  {author} {\bibinfo {author} {\bibfnamefont {M.}~\bibnamefont
  {Valiente}}\ and\ \bibinfo {author} {\bibfnamefont {D.}~\bibnamefont
  {Petrosyan}},\ }\href {http://stacks.iop.org/0953-4075/41/i=16/a=161002}
  {\bibfield  {journal} {\bibinfo  {journal} {J. Phys. B}\ }\textbf {\bibinfo
  {volume} {41}},\ \bibinfo {pages} {161002} (\bibinfo {year}
  {2008})}\BibitemShut {NoStop}%
\bibitem [{\citenamefont {Javanainen}\ \emph {et~al.}(2010)\citenamefont
  {Javanainen}, \citenamefont {Odong},\ and\ \citenamefont
  {Sanders}}]{Javanainen2010}%
  \BibitemOpen
  \bibfield  {author} {\bibinfo {author} {\bibfnamefont {J.}~\bibnamefont
  {Javanainen}}, \bibinfo {author} {\bibfnamefont {O.}~\bibnamefont {Odong}}, \
  and\ \bibinfo {author} {\bibfnamefont {J.~C.}\ \bibnamefont {Sanders}},\
  }\href {\doibase 10.1103/PhysRevA.81.043609} {\bibfield  {journal} {\bibinfo
  {journal} {Phys. Rev. A}\ }\textbf {\bibinfo {volume} {81}},\ \bibinfo
  {pages} {043609} (\bibinfo {year} {2010})}\BibitemShut {NoStop}%
\bibitem [{\citenamefont {Wang}\ and\ \citenamefont {Liang}(2010)}]{Wang2010}%
  \BibitemOpen
  \bibfield  {author} {\bibinfo {author} {\bibfnamefont {Y.-M.}\ \bibnamefont
  {Wang}}\ and\ \bibinfo {author} {\bibfnamefont {J.-Q.}\ \bibnamefont
  {Liang}},\ }\href {\doibase 10.1103/PhysRevA.81.045601} {\bibfield  {journal}
  {\bibinfo  {journal} {Phys. Rev. A}\ }\textbf {\bibinfo {volume} {81}},\
  \bibinfo {pages} {045601} (\bibinfo {year} {2010})}\BibitemShut {NoStop}%
\bibitem [{\citenamefont {Nguenang}\ \emph {et~al.}(2012)\citenamefont
  {Nguenang}, \citenamefont {Flach},\ and\ \citenamefont
  {Khomeriki}}]{Nguenang2012}%
  \BibitemOpen
  \bibfield  {author} {\bibinfo {author} {\bibfnamefont {J.-P.}\ \bibnamefont
  {Nguenang}}, \bibinfo {author} {\bibfnamefont {S.}~\bibnamefont {Flach}}, \
  and\ \bibinfo {author} {\bibfnamefont {R.}~\bibnamefont {Khomeriki}},\ }\href
  {\doibase http://dx.doi.org/10.1016/j.physleta.2011.11.048} {\bibfield
  {journal} {\bibinfo  {journal} {Phys. Lett. A}\ }\textbf {\bibinfo {volume}
  {376}},\ \bibinfo {pages} {472 } (\bibinfo {year} {2012})}\BibitemShut
  {NoStop}%
\bibitem [{\citenamefont {Qin}\ \emph {et~al.}(2014)\citenamefont {Qin},
  \citenamefont {Ke}, \citenamefont {Guan}, \citenamefont {Li}, \citenamefont
  {Andrei},\ and\ \citenamefont {Lee}}]{Qin2014}%
  \BibitemOpen
  \bibfield  {author} {\bibinfo {author} {\bibfnamefont {X.}~\bibnamefont
  {Qin}}, \bibinfo {author} {\bibfnamefont {Y.}~\bibnamefont {Ke}}, \bibinfo
  {author} {\bibfnamefont {X.}~\bibnamefont {Guan}}, \bibinfo {author}
  {\bibfnamefont {Z.}~\bibnamefont {Li}}, \bibinfo {author} {\bibfnamefont
  {N.}~\bibnamefont {Andrei}}, \ and\ \bibinfo {author} {\bibfnamefont
  {C.}~\bibnamefont {Lee}},\ }\href {\doibase 10.1103/PhysRevA.90.062301}
  {\bibfield  {journal} {\bibinfo  {journal} {Phys. Rev. A}\ }\textbf {\bibinfo
  {volume} {90}},\ \bibinfo {pages} {062301} (\bibinfo {year}
  {2014})}\BibitemShut {NoStop}%
\bibitem [{\citenamefont {Bello}\ \emph {et~al.}(2016)\citenamefont {Bello},
  \citenamefont {Creffield},\ and\ \citenamefont {Platero}}]{Bello2016}%
  \BibitemOpen
  \bibfield  {author} {\bibinfo {author} {\bibfnamefont {M.}~\bibnamefont
  {Bello}}, \bibinfo {author} {\bibfnamefont {C.~E.}\ \bibnamefont
  {Creffield}}, \ and\ \bibinfo {author} {\bibfnamefont {G.}~\bibnamefont
  {Platero}},\ }\href {http://dx.doi.org/10.1038/srep22562} {\bibfield
  {journal} {\bibinfo  {journal} {Sci. Rep.}\ }\textbf {\bibinfo {volume}
  {6}},\ \bibinfo {pages} {22562} (\bibinfo {year} {2016})}\BibitemShut
  {NoStop}%
\bibitem [{\citenamefont {Di~Liberto}\ \emph {et~al.}(2016)\citenamefont
  {Di~Liberto}, \citenamefont {Recati}, \citenamefont {Carusotto},\ and\
  \citenamefont {Menotti}}]{Liberto2016}%
  \BibitemOpen
  \bibfield  {author} {\bibinfo {author} {\bibfnamefont {M.}~\bibnamefont
  {Di~Liberto}}, \bibinfo {author} {\bibfnamefont {A.}~\bibnamefont {Recati}},
  \bibinfo {author} {\bibfnamefont {I.}~\bibnamefont {Carusotto}}, \ and\
  \bibinfo {author} {\bibfnamefont {C.}~\bibnamefont {Menotti}},\ }\href
  {\doibase 10.1103/PhysRevA.94.062704} {\bibfield  {journal} {\bibinfo
  {journal} {Phys. Rev. A}\ }\textbf {\bibinfo {volume} {94}},\ \bibinfo
  {pages} {062704} (\bibinfo {year} {2016})}\BibitemShut {NoStop}%
\bibitem [{\citenamefont {{Gorlach}}\ and\ \citenamefont
  {{Poddubny}}(2016)}]{Gorlach2016}%
  \BibitemOpen
  \bibfield  {author} {\bibinfo {author} {\bibfnamefont {M.~A.}\ \bibnamefont
  {{Gorlach}}}\ and\ \bibinfo {author} {\bibfnamefont {A.~N.}\ \bibnamefont
  {{Poddubny}}},\ }\href@noop {} {\bibfield  {journal} {\bibinfo  {journal}
  {ArXiv e-prints}\ } (\bibinfo {year} {2016})},\ \Eprint
  {http://arxiv.org/abs/1608.02093} {arXiv:1608.02093 [cond-mat.mes-hall]}
  \BibitemShut {NoStop}%
\bibitem [{\citenamefont {Wada}(1992)}]{Wada1992}%
  \BibitemOpen
  \bibfield  {author} {\bibinfo {author} {\bibfnamefont {Y.}~\bibnamefont
  {Wada}},\ }\enquote {\bibinfo {title} {New horizons in low-dimensional
  electron systems: A festschrift in honour of professor h. kamimura},}\ \
  (\bibinfo  {publisher} {Springer Netherlands},\ \bibinfo {address}
  {Dordrecht},\ \bibinfo {year} {1992})\ Chap.\ \bibinfo {chapter} {Doping and
  Disorder in Conducting Polymers}, pp.\ \bibinfo {pages}
  {415--432}\BibitemShut {NoStop}%
\bibitem [{\citenamefont {Hirsch}\ \emph {et~al.}(1982)\citenamefont {Hirsch},
  \citenamefont {Sugar}, \citenamefont {Scalapino},\ and\ \citenamefont
  {Blankenbecler}}]{Hirsch1982}%
  \BibitemOpen
  \bibfield  {author} {\bibinfo {author} {\bibfnamefont {J.~E.}\ \bibnamefont
  {Hirsch}}, \bibinfo {author} {\bibfnamefont {R.~L.}\ \bibnamefont {Sugar}},
  \bibinfo {author} {\bibfnamefont {D.~J.}\ \bibnamefont {Scalapino}}, \ and\
  \bibinfo {author} {\bibfnamefont {R.}~\bibnamefont {Blankenbecler}},\ }\href
  {\doibase 10.1103/PhysRevB.26.5033} {\bibfield  {journal} {\bibinfo
  {journal} {Phys. Rev. B}\ }\textbf {\bibinfo {volume} {26}},\ \bibinfo
  {pages} {5033} (\bibinfo {year} {1982})}\BibitemShut {NoStop}%
\bibitem [{\citenamefont {Cannon}\ \emph {et~al.}(1991)\citenamefont {Cannon},
  \citenamefont {Scalettar},\ and\ \citenamefont {Fradkin}}]{Cannon1991}%
  \BibitemOpen
  \bibfield  {author} {\bibinfo {author} {\bibfnamefont {J.~W.}\ \bibnamefont
  {Cannon}}, \bibinfo {author} {\bibfnamefont {R.~T.}\ \bibnamefont
  {Scalettar}}, \ and\ \bibinfo {author} {\bibfnamefont {E.}~\bibnamefont
  {Fradkin}},\ }\href {\doibase 10.1103/PhysRevB.44.5995} {\bibfield  {journal}
  {\bibinfo  {journal} {Phys. Rev. B}\ }\textbf {\bibinfo {volume} {44}},\
  \bibinfo {pages} {5995} (\bibinfo {year} {1991})}\BibitemShut {NoStop}%
\bibitem [{\citenamefont {Voit}(1992)}]{Voit1992}%
  \BibitemOpen
  \bibfield  {author} {\bibinfo {author} {\bibfnamefont {J.}~\bibnamefont
  {Voit}},\ }\href {\doibase 10.1103/PhysRevB.45.4027} {\bibfield  {journal}
  {\bibinfo  {journal} {Phys. Rev. B}\ }\textbf {\bibinfo {volume} {45}},\
  \bibinfo {pages} {4027} (\bibinfo {year} {1992})}\BibitemShut {NoStop}%
\bibitem [{\citenamefont {Mila}\ and\ \citenamefont {Zotos}(1993)}]{Mila1993}%
  \BibitemOpen
  \bibfield  {author} {\bibinfo {author} {\bibfnamefont {F.}~\bibnamefont
  {Mila}}\ and\ \bibinfo {author} {\bibfnamefont {X.}~\bibnamefont {Zotos}},\
  }\href {http://stacks.iop.org/0295-5075/24/i=2/a=010} {\bibfield  {journal}
  {\bibinfo  {journal} {EPL}\ }\textbf {\bibinfo {volume} {24}},\ \bibinfo
  {pages} {133} (\bibinfo {year} {1993})}\BibitemShut {NoStop}%
\bibitem [{\citenamefont {van Dongen}(1994)}]{Dongen1994}%
  \BibitemOpen
  \bibfield  {author} {\bibinfo {author} {\bibfnamefont {P.~G.~J.}\
  \bibnamefont {van Dongen}},\ }\href {\doibase 10.1103/PhysRevB.49.7904}
  {\bibfield  {journal} {\bibinfo  {journal} {Phys. Rev. B}\ }\textbf {\bibinfo
  {volume} {49}},\ \bibinfo {pages} {7904} (\bibinfo {year}
  {1994})}\BibitemShut {NoStop}%
\bibitem [{\citenamefont {Penc}\ and\ \citenamefont {Mila}(1994)}]{Penc1994}%
  \BibitemOpen
  \bibfield  {author} {\bibinfo {author} {\bibfnamefont {K.}~\bibnamefont
  {Penc}}\ and\ \bibinfo {author} {\bibfnamefont {F.}~\bibnamefont {Mila}},\
  }\href {\doibase 10.1103/PhysRevB.49.9670} {\bibfield  {journal} {\bibinfo
  {journal} {Phys. Rev. B}\ }\textbf {\bibinfo {volume} {49}},\ \bibinfo
  {pages} {9670} (\bibinfo {year} {1994})}\BibitemShut {NoStop}%
\bibitem [{\citenamefont {Ejima}\ \emph {et~al.}(2016)\citenamefont {Ejima},
  \citenamefont {Essler}, \citenamefont {Lange},\ and\ \citenamefont
  {Fehske}}]{Ejima2016}%
  \BibitemOpen
  \bibfield  {author} {\bibinfo {author} {\bibfnamefont {S.}~\bibnamefont
  {Ejima}}, \bibinfo {author} {\bibfnamefont {F.~H.~L.}\ \bibnamefont
  {Essler}}, \bibinfo {author} {\bibfnamefont {F.}~\bibnamefont {Lange}}, \
  and\ \bibinfo {author} {\bibfnamefont {H.}~\bibnamefont {Fehske}},\ }\href
  {\doibase 10.1103/PhysRevB.93.235118} {\bibfield  {journal} {\bibinfo
  {journal} {Phys. Rev. B}\ }\textbf {\bibinfo {volume} {93}},\ \bibinfo
  {pages} {235118} (\bibinfo {year} {2016})}\BibitemShut {NoStop}%
\bibitem [{\citenamefont {Orbach}(1958)}]{Orbach1958}%
  \BibitemOpen
  \bibfield  {author} {\bibinfo {author} {\bibfnamefont {R.}~\bibnamefont
  {Orbach}},\ }\href {\doibase 10.1103/PhysRev.112.309} {\bibfield  {journal}
  {\bibinfo  {journal} {Phys. Rev.}\ }\textbf {\bibinfo {volume} {112}},\
  \bibinfo {pages} {309} (\bibinfo {year} {1958})}\BibitemShut {NoStop}%
\bibitem [{\citenamefont {Lacroix}\ \emph {et~al.}(2011)\citenamefont
  {Lacroix}, \citenamefont {Mendels},\ and\ \citenamefont
  {Mila}}]{Lacroix2011}%
  \BibitemOpen
  \bibfield  {author} {\bibinfo {author} {\bibfnamefont {C.}~\bibnamefont
  {Lacroix}}, \bibinfo {author} {\bibfnamefont {P.}~\bibnamefont {Mendels}}, \
  and\ \bibinfo {author} {\bibfnamefont {F.}~\bibnamefont {Mila}},\ }\href@noop
  {} {\emph {\bibinfo {title} {Introduction to Frustrated Magnetism: Materials,
  Experiments, Theory}}},\ edited by\ \bibinfo {editor} {\bibnamefont
  {Springer}}\ (\bibinfo  {publisher} {Springer Series in Solid-State
  Sciences},\ \bibinfo {year} {2011})\BibitemShut {NoStop}%
\end{thebibliography}%

\end{document}